\documentclass[aps,prl,preprint,groupedaddress]{revtex4-1}

\bibstyle{apsrev4-1}

\usepackage{graphicx}
\usepackage{amssymb}
\usepackage{epstopdf}
\begin{document}

\preprint{}

\title{Statistical Physics of Adaptation}


\author{Nikolai Perunov, Robert Marsland, and Jeremy England}
\affiliation{Department of Physics, Physics of Living Systems Group, Massachusetts Institute of Technology, Floor 6, 400 Tech Square, Cambridge, MA 02139}


\date{\today}

\begin{abstract}
All living things exhibit adaptations that enable them to survive and reproduce in the natural environment that they inhabit.  From a biological standpoint, it has long been understood that adaptation comes from natural selection, whereby maladapted individuals do not pass their traits effectively to future generations.  However, we may also consider the phenomenon of adaptation from the standpoint of physics, and ask whether it is possible to delineate what the difference is in terms of physical properties between something that is well-adapted to its surrounding environment, and something that is not.  In this work, we undertake to address this question from a theoretical standpoint.  Building on past fundamental results in far-from-equilibrium statistical mechanics, we demonstrate a generalization of the Helmholtz free energy for the finite-time stochastic evolution of driven Newtonian matter.  By analyzing this expression term by term, we are able to argue for a general tendency in driven many-particle systems towards self-organization into states formed through exceptionally reliable absorption and dissipation of work energy from the surrounding environment.  Subsequently, we illustrate the mechanism of this general tendency towards physical adaptation by analyzing the process of random hopping in driven energy landscapes.  
\end{abstract}

\pacs{}

\maketitle


\maketitle

\section{Introduction}

What does it mean to say a thing is well-adapted to its environment?  Of course, any succinct answer we might try give to this question would fall short of addressing the whole issue, yet we can still readily appreciate that the easiest way forward lies in invoking the biological language of Darwinian evolution: an organism living in a particular environment is well-adapted if its qualities, behaviors, and capacities generally enable it to survive and reproduce.  The substantial explanatory power of this analytical frame can best be appreciated by considering idealized examples: if two microbes in a lab experiment are nearly identical, but one of them self-replicates slightly faster due to the elevated catalytic rate of some particular enzyme, we expect the faster grower's progeny (possessing genomes encoding the faster enzyme)  to take over the flask; meanwhile, at the opposite extreme, common sense tells us that a cow submerged in deep ocean water will not last long enough to calve.  In both cases, adaptation can be practically defined in the negative by saying that a species that is insufficiently adapted to its environment is one that will not persist there.

One difficulty with thinking about adaptation in this way is that its account of what adaptive success looks like risks being too vague; both blue whales and phytoplankton persist and propagate in the same ocean, yet they do so by mechanisms that are at once both dramatically distinct and strongly intertwined.  Clearly, when comparing such different organisms to each other, it is meaningless to ask which is more ``fit" under a single set of environmental conditions in which they flourish interdependently.  Nonetheless, our study of evolution on long timescales fills us with the striking sense of directional motion and progress -- even in complexly interacting ensembles of organisms assaulted by constant environmental fluctuations -- which raises the question of whether we might find a language in which the character of this progress could be more rigorously described.

The first step towards developing such a language is to recognize a problem of categorization in the way we have defined the problem.  The autonomy and behavioral dynamism of individual organisms is so impressive to us that, when thinking like biologists, we tend to conceive of the single living creature  as the fundamental unit through which all other phenomena (be they biochemical or ecological) should be understood.  Of course, we may complicate this view by instead recommending that single cells or genes be taken as the focus \cite{dawkins2006selfish}.  Still, from the theoretical standpoint of Newtonian physics, whales, algal blooms, and bits of DNA are all just different ways of clumping together the same material building blocks by assigning position and momentum coordinates to each atom.  
Viewed in this light, we may entertain the notion that it is somewhat arbitrary how we have decided to draw a line around a whale, or around a segment of chromosome, and say that this is one indivisible whole possessing a separate identity from the rest of the matter in the system.  Even more intriguingly, we are justified in asking whether it is possible to construct a new notion of how ``adapted" the clumps of matter we observe are to their environment -- one that can be expressed purely in terms of their physical properties, without needing to decide in advance on a physical definition of where one replicator ends, and another begins.  

The prospect of developing a clear definition for ``physical adaptation" is particularly appealing when we consider that a wide variety of physical systems have nothing alive in them, but nonetheless exhibit dynamics rather reminiscent of biological evolution.  Just as whales and algae are structures that emerged on a planet driven far from thermal equilibrium by the light of the sun, so too do intricately organized structures appear in a variety of other far-from-equilibrium systems \cite{prigogine1971biological}.  And just as one might say of sea creatures that their highly adapted forms clearly reflect various properties of the environment in which they evolved, so too may one say of a broad class of hysteretic nonequilibrium systems that, as they evolve stochastically, the structures that emerge in them accumulate information about the external drives to which they have been subject.  However, while in biology, Darwinian concepts of fitness and selection provide a means to explain why some structures persist and others do not in a given environment, the question of whether a more general account can be given from the standpoint of physics in self-organized nonequilibrium systems remains tantalizingly open.

In this article, our aim to is to provide the beginnings of such an account.   Recent theoretical progress in nonequilibrium statistical mechanics \cite{crooks, England2013} has established fundamental relationships that hold between the rates of stochastic irreversible transitions in driven open systems and the amounts of entropy production that must accompany these transitions.  Using these results, we will derive and analyze a generalization of the Helmholtz free energy for out-of-equilibrium macroscopic systems, arguing that driven stochastic evolution can favor the discovery of organized states that form through increased dissipation and the suppression of fluctuations.  Subsequently, we will demonstrate the mechanism of this general relationship in simple analytical examples, and discuss the relevance of our findings to the study of evolution and self-organization.

\section{Entropy Production and Stochastic Evolution}
The general scenario in which adaptive self-organization might be explored is one where some large number of particles (one could say: a certain number of kilograms of carbon, nitrogen, oxygen, etc.) are confined to a volume $V$ that is held in contact with a heat bath at constant inverse temperature $\beta\equiv 1/T$.  In accordance with the approach pioneered in previous works \cite{jarzynski1997nonequilibrium}, the classical Hamiltonian of the whole set-up is written as
\begin{equation}
H_{tot} = H_{sys}(\mathbf{x},\lambda(t))+H_{bath}(\mathbf{y})+h_{int}(\mathbf{x},\mathbf{y})
\end{equation}
where $\mathbf{x} = \{q^{(1)}_{1},p^{(1)}_{1},\ldots,q^{(3)}_{N},p^{(3)}_{N}\}$ accounts for all the coordinate degrees of freedom for the $N$ particles in the system, $\mathbf{y}$ does the same for the bath, and the Hamiltonian functions $H_{sys}$, $H_{bath}$, and $h_{int}$ define conservative interactions among the various position coordinates of system and bath.  The function $\lambda(t)$ plays the role of a time-varying external field that acts exclusively on the \emph{system} and can do work on the coordinates $\mathbf{x}$. Crucially, $h_{int}$ is assumed to be small, so that the term merely plays the formal role of enabling the flow of energy between the bath and the system, and may otherwise be ignored.     

The standard approach \cite{crooks} to modeling such a system is to posit that the coupling to the heat bath introduces stochasticity into the observed dynamics, so that if one starts off at some particular point in the system's phase space $\mathbf{x}(0)$ at time $t=0$, then, for a given choice of the driving field $\lambda(t)$, there is some probability density for microtrajectories of the system $\pi_{\tau}[\mathbf{x}(t)| \mathbf{x}(0);\lambda(t)]$ that expresses how likely one would be to observe the system to progress through a given series of subsequent arrangements $\mathbf{x}(t)$ over time $\tau$.

It is worth stressing at the outset that the space of distinct possible arrangements for the system described above is expected to be mind-bogglingly vast, and inconceivably diverse.  Considering, for example, the number of different ways of assigning positions and momenta to all of the different atoms that make up a blue whale, one should expect to find in the phase space for such a system other arrangements of the particles that look nothing like a whale, and indeed could instead look like a whole school of smaller fish, or a wide assortment of furniture pieces.  This point is worth emphasizing because it means that, in principle, the functional $\pi_{\tau}[\mathbf{x}(t)| \mathbf{x}(0);\lambda(t)]$ should be well-defined for any time-sequence of microstates, no matter how outlandish: it may be very unlikely indeed to see a blue whale spontaneously turn into a bunch of fish in the space of five minutes, but formally we are required to say in this modeling framework that the classical probability of such an occurrence taking place as the result of a freak thermal fluctuation is greater than zero.  

Laughable as it is to make this point, doing so provokes us to ask the right question, namely: what is it that makes some stochastic trajectories for the system more likely than others?  For the beginnings of an answer, we should first look to the microscopic reversibility relation put forward by Crooks \cite{crooks}:
\begin{equation}
\frac{\pi_{\tau}[\mathbf{x}^{*}_{\tau}(\tau-t)| \mathbf{x}^{*}(\tau);\lambda(\tau-t)]}{\pi_{\tau}[\mathbf{x}(t)| \mathbf{x}(0);\lambda(t)]}=\exp\left(-\beta \Delta Q[\mathbf{x}(t)]\right)=\exp\left(-\Delta S_{bath}[\mathbf{x}(t)]\right)
\end{equation}
This equation relates the probability of observing a given microtrajectory to the probability of observing the time-reversed movie trajectory after flipping all momentum coordinates in the starting state ($\mathbf{x}^{*}(\tau) = (q^{(1)}_{1}(\tau),-p^{(1)}_{1}(\tau),\ldots)$).  Because of the underlying time-reversal symmetry of Newton's laws, the ratio of the forward and reverse probabilities must be exactly equal to the exponential of the heat $\beta\Delta Q[\mathbf{x}(t)]$ evolved into the bath in the forward direction as the system traverses the microtrajectory $\mathbf{x}(t)$.  Since this heat is being transferred to a vast external reservoir at constant temperature $T=1/\beta$, we may accordingly identify it thermodynamically as entropy production and write $\beta \Delta Q[\mathbf{x}(t)] = \Delta S_{bath}[\mathbf{x}(t)] $.  Thus, the Crooks result establishes a general, exact, microscopically-detailed relationship between statistical irreversibility and entropy production in the surrounding bath, and for the remainder of this work we will treat heat evolution and entropy production interchangeably.

Recently \cite{England2013}, we have demonstrated the consequences of the above relation for the stochastic dynamics of arbitrarily driven nonequilibrium macrostates.  We may define a macrostate by labeling an arbitrary collection of microstates as sharing some macro-observable property $\mathbf{I}$, and in that case, if the system is prepared according to some controlled experimental procedure to be in $\mathbf{I}$, then implicitly it will possess some nonequilibrium density over microstates $p(\mathbf{x}|\mathbf{I})~d\mathbf{x}\equiv p_{i}(\mathbf{x})~d\mathbf{x}$.  If we furthermore designate some other disjoint set of microstates to have macro-property $\mathbf{II}$, then we can define $\pi_{\tau}[\mathbf{I}\rightarrow\mathbf{II};\lambda(t)]$ as the probability that the system is observed to have property $\mathbf{II}$ after stochastically evolving under the applied field $\lambda(t)$ for time $\tau$.  And if the system is indeed observed to be in $\mathbf{II}$, then implicitly it will be distributed over the microstates available to it according to some new density $p(\mathbf{x}|\mathbf{II},\mathbf{I};\lambda(t);\tau)\equiv p_{f}(\mathbf{x})~d\mathbf{x}$.  In this case, we can also define the probability $\pi^{\mathbf{I}}_{\tau}[\mathbf{II}^{*}\rightarrow\mathbf{I}^{*};\lambda(\tau-t)]$ of reverting back to $\mathbf{I}$ if particle momenta are reversed and the applied field is run backward (Figure 1).  With these definitions in hand, it may be shown that
\begin{equation}
\frac{\pi^{\mathbf{I}}_{\tau}[\mathbf{II}^{*}\rightarrow\mathbf{I}^{*};\lambda(\tau-t)]}{\pi_{\tau}[\mathbf{I}\rightarrow\mathbf{II};\lambda(t)]} = \big\langle\exp(-\Delta S_{tot})\big\rangle_{\mathbf{I}\rightarrow\mathbf{II}}
\end{equation}
where $\Delta S_{tot}[\mathbf{x}(t)] \equiv \ln\left[\frac{p_{i}(\mathbf{x}(0))}{p_{f}(\mathbf{x}(\tau))}\right]+\beta\Delta Q[\mathbf{x}(t)]$.  Put succinctly, this result says that statistical irreversibility measured on arbitrary macro-observables still has an exact quantitative relationship to entropy production, with two important differences: firstly, that one must account for the internal entropy difference between macrostates, and secondly that now the entropy must be exponentially averaged in the appropriate way over the statistical weights of all the microtrajectories propagating from the starting macrostate to the ending macrostate.

 \begin{figure}
 \includegraphics[width=\columnwidth]{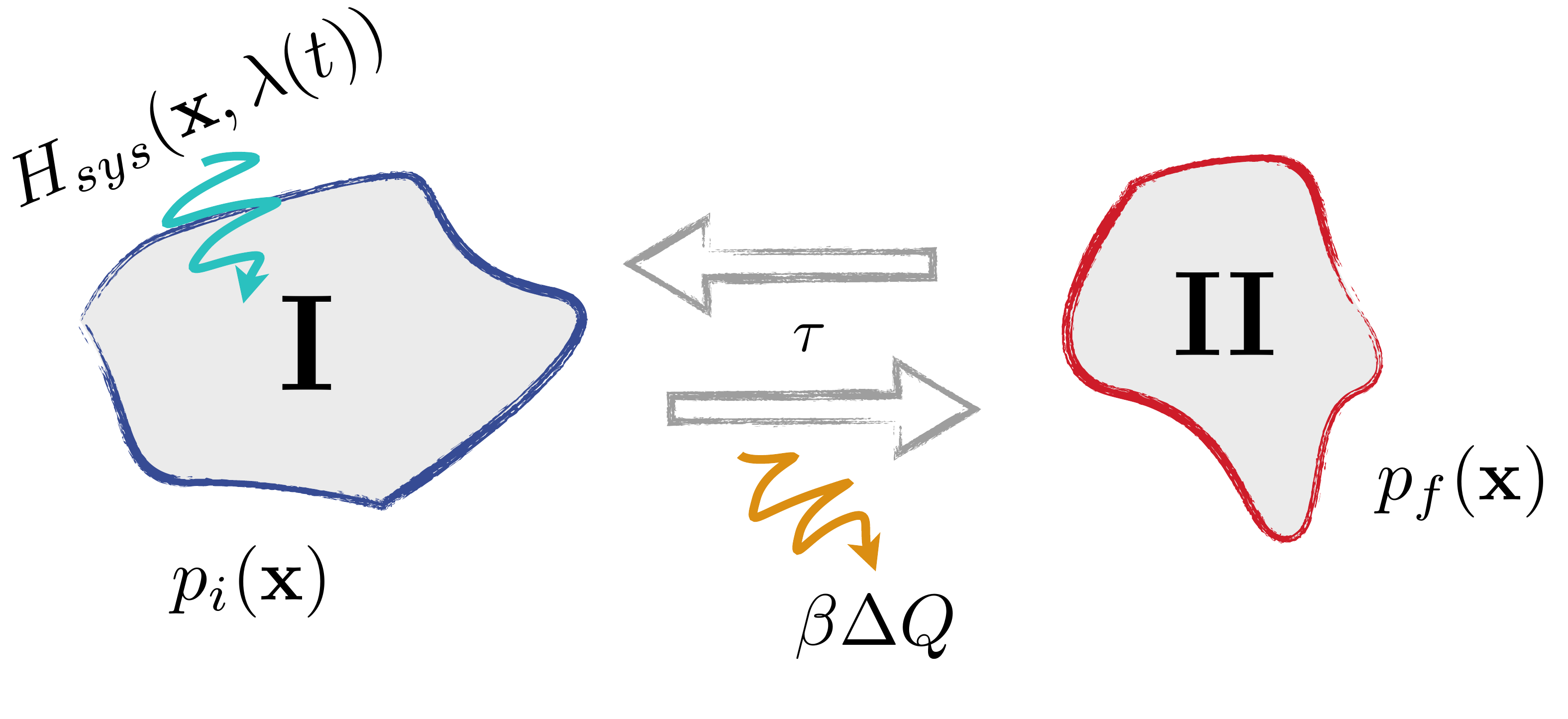}
 \caption{An arbitrary nonequilibrium macrostate for the system of interest is constructed as the probability density $p_{i}(\mathbf{x})~d\mathbf{x}$ over the microstates $\mathbf{x}$ when the system is prepared by a controlled procedure to have some macroscopic property $\mathbf{I}$.  The system is externally driven for period of duration $\tau$ by a time-varying field $H_{sys}(\mathbf{x},\lambda(t))$ while in contact with a thermal bath of inverse temperature $\beta$ that absorbs heat $\Delta Q$.  Afterwards, one may ask whether the system now has a new macroscopic property $\mathbf{II}$; if so, the new probability distribution over microstates is implicitly given by $p_{f}(\mathbf{x})~d\mathbf{x}$.  }
 \end{figure}

Previously, the above result was used to derive a generalization of the Second Law of Thermodynamics, which was applied to studying the thermodynamic constraints obeyed by self-replicators as they grow.  Here, we will instead use the equation to give a thermodynamic account of driven stochastic evolution.  Thus, let us consider a many-particle system prepared in $\mathbf{I}$ in the manner described above, and let us suppose that we have defined two distinct possible macroscopic outcomes, $\mathbf{II}$ and $\mathbf{III}$ that could lie in the system's future (Figure 2).  A question of great interest to us would be which of the two outcomes should be more likely given $H_{sys}(\mathbf{x},\lambda(t))$, that is, given the way the particles inside the system interact with each other and the way the system is being externally driven.   Formally, we may write
\begin{equation}
\ln\left[\frac{\pi_{\tau}[\mathbf{I}\rightarrow\mathbf{II};\lambda(t)]}{\pi_{\tau}[\mathbf{I}\rightarrow\mathbf{III};\lambda(t)]}\right]=\ln\left[\frac{\pi^{\mathbf{I}}_{\tau}[\mathbf{II}^{*}\rightarrow\mathbf{I}^{*};\lambda(\tau-t)]}{\pi^{\mathbf{I}}_{\tau}[\mathbf{III}^{*}\rightarrow\mathbf{I}^{*};\lambda(\tau-t)]}\right]-\ln\left[\frac{ \big\langle\exp(-\Delta S_{tot})\big\rangle_{\mathbf{I}\rightarrow\mathbf{II}}}{ \big\langle\exp(-\Delta S_{tot})\big\rangle_{\mathbf{I}\rightarrow\mathbf{III}}}\right]
\end{equation}
One can already say a great deal from the way this equation is set down, but a few additional simplifying assumptions help to make its structure even clearer: if we assume that the system is driven for a long time, it may be reasonable to posit that there is no correlation between $\mathbf{x}(0)$ and $\mathbf{x}(\tau)$ other than the one introduced by constraining the starting and ending points to lie in certain macrostates. And if we moreover assume that $p_{i}$ and $p_{f}$ are close enough to uniform over their respective macrostates that $1/p\sim \Omega$ provides a good measure of the phase space volume taken up by each macrostate, then we obtain
\begin{equation}
\ln\left[\frac{\pi_{\tau}[\mathbf{I}\rightarrow\mathbf{II};\lambda(t)]}{\pi_{\tau}[\mathbf{I}\rightarrow\mathbf{III};\lambda(t)]}\right]=\Delta\ln\Omega_{\mathbf{II},\mathbf{III}}+\ln\left[\frac{\pi^{\mathbf{I}}_{\tau}[\mathbf{II}^{*}\rightarrow\mathbf{I}^{*};\lambda(\tau-t)]}{\pi^{\mathbf{I}}_{\tau}[\mathbf{III}^{*}\rightarrow\mathbf{I}^{*};\lambda(\tau-t)]}\right]-\ln\left[\frac{ \big\langle\exp(-\beta\Delta Q)\big\rangle_{\mathbf{I}\rightarrow\mathbf{II}}}{ \big\langle\exp(-\beta\Delta Q)\big\rangle_{\mathbf{I}\rightarrow\mathbf{III}}}\right]
\end{equation}
While the above thermodynamic expression may be correct, there are two significant reasons to doubt that it could be at all useful.  The first of these is that we are purportedly interested in knowing how much more likely we are to see $\mathbf{II}$ happen in our driven system than $\mathbf{III}$, and yet in order to say anything concrete it would seem that we have to know the reversal probabilities $\pi^{\mathbf{I}}_{\tau}$, which requires complete knowledge of all the rates of stochastic transitions that are possible in our system.  Put another way, we may as well run a fully-detailed microscopic simulation and just see what happens.  The second problem is the well-known difficulty of averaging exponentials of large quantities: there are many microtrajectories that run from $\mathbf{I}$ to $\mathbf{II}$, and in general it will be some sub-ensemble of highly improbable ones that make the dominant contribution to $\big\langle\exp(-\beta\Delta Q)\big\rangle_{\mathbf{I}\rightarrow\mathbf{II}}$.  This puts us in the untenable position of needing to know the heat dissipation associated with stochastic events in our system that are immeasurably unlikely to occur \cite{jarzynski2006rare}.

 \begin{figure}
 \includegraphics[width=\columnwidth]{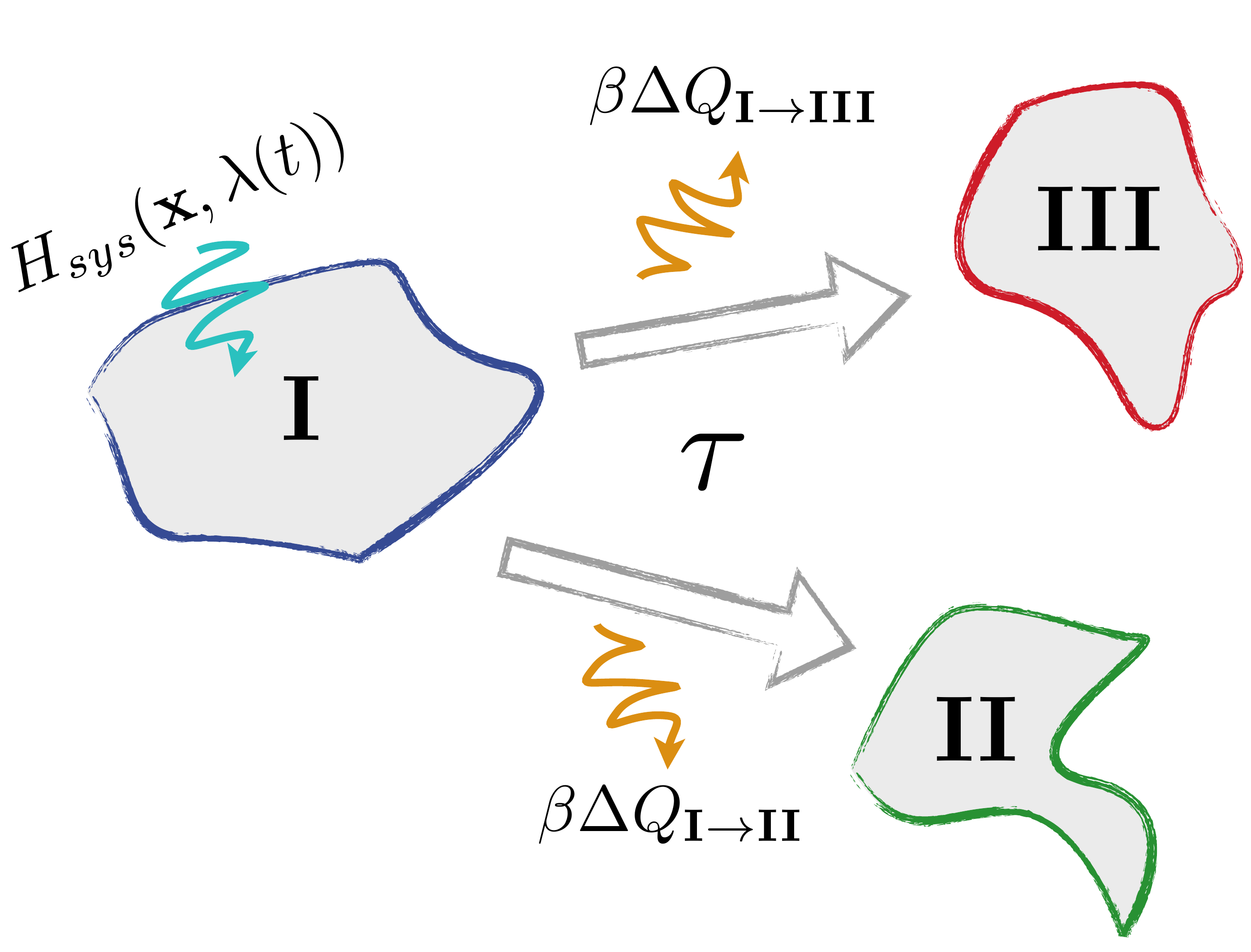}
 \caption{In this scenario, we consider a starting nonequilibrium macrostate $\mathbf{I}$ that is driven by an external field while it evolves stochastically in contact with a thermal bath.  We may ask whether, after some finite time $\tau$ it is more likely to be found in one of two other possible macrostates $\mathbf{II}$ and $\mathbf{III}$.  The relative likelihood of these two different possibilities will be determined in part by the statistical distributions for possible values of $\beta\Delta Q_{\mathbf{I}\rightarrow *}$ in each case.  }
 \end{figure}

We can start at least by turning two problems into one if we note that everything we have said so far still applies if we choose not to average over \emph{all} paths leading from one macrostate to another, but rather only over a select sub-ensemble of paths containing most of the probability current.  If we perform this average over likely forward paths, the two things that change in tandem are the dissipated heat averages and the reversal probabilities: we get to average heat over events of measurable likelihood, and the ``price" we pay is that the reversal probabilities now correspond only to the likelihoods of trajectories that look like reverse movies of trajectories that are likely in the forward direction.  We can denote this more compactly via
\begin{equation}
\ln\left[\frac{\pi^{\mathbf{I},\mathbf{II},fwd}_{\tau,\lambda(t)}}{\pi^{\mathbf{I},\mathbf{III},fwd}_{\tau,\lambda(t)}}\right]=\Delta\ln\Omega_{\mathbf{II},\mathbf{III}}+\ln\left[\frac{\pi^{\mathbf{I},\mathbf{II},rev}_{\tau,\lambda(t)}}{\pi^{\mathbf{I},\mathbf{III},rev}_{\tau,\lambda(t)}}\right]-\ln\left[\frac{ \big\langle\exp(-\beta\Delta Q)\big\rangle^{fwd}_{\mathbf{I}\rightarrow\mathbf{II}}}{ \big\langle\exp(-\beta\Delta Q)\big\rangle^{fwd}_{\mathbf{I}\rightarrow\mathbf{III}}}\right]
\end{equation}

Taking one final heuristic step, we may also note that $-\ln\langle\exp(-\beta \Delta Q)\rangle$ is a cumulant generating function when we expand in $\beta$, and thus may be written as
\begin{equation}
-\ln\langle\exp(-\beta \Delta Q)\rangle = \beta\langle \Delta Q\rangle-\frac{\beta^{2}}{2}\sigma_{\Delta Q}^{2}+\ldots \equiv \Psi -\Phi
\end{equation}
where $\Psi = \beta\langle \Delta Q\rangle$ is the average dissipated heat and $\Phi$ is accordingly defined to account for all of the fluctuations about this average via $\Phi =\ln\langle\exp(-\beta \Delta Q)\rangle+\Psi$. Thus, we may write
\begin{equation}
\label{eq:boltz}
\ln\left[\frac{\pi^{\mathbf{I},\mathbf{II},fwd}_{\tau,\lambda(t)}}{\pi^{\mathbf{I},\mathbf{III},fwd}_{\tau,\lambda(t)}}\right]=\Delta\ln\Omega_{\mathbf{II},\mathbf{III}}+\ln\left[\frac{\pi^{\mathbf{I},\mathbf{II},rev}_{\tau,\lambda(t)}}{\pi^{\mathbf{I},\mathbf{III},rev}_{\tau,\lambda(t)}}\right]+\Delta\Psi^{\mathbf{I},fwd}_{\mathbf{II},\mathbf{III}}-\Delta\Phi^{\mathbf{I},fwd}_{\mathbf{II},\mathbf{III}}
\end{equation}
Equation (\ref{eq:boltz}) has a rich structure, and may be thought of as a generalization of the Helmholtz free energy for the stochastic evolution of arbitrarily driven nonequilibrium macrostates over finite time.  To understand it, it is most helpful to consider how the left-hand side is affected when one term on the right-hand side is varied while holding the others fixed. The first term has the most obvious and familiar meaning, because it is identical to what we would see in the undriven equilibrium expression for free energy: all things being equal, an outcome is more likely if it has higher internal entropy ($\ln\Omega)$, that is, if it is a macrostate that corresponds to a larger volume in the system's phase space.

The second term has no analog in equilibrium thermodynamics and is inherently kinetic in nature.  Succinctly, it says that, in a given amount of time, one is more likely to propagate to a macrostate from which one is more likely to revert back in the same amount of time.  The most intuitive way of understanding the effect of this term is to consider a simple transition-state theory model of a one-step chemical reaction \cite{gardiner}: when the activation free energy barrier is lowered, both the forward and reverse rates of the reaction are accelerated.  Thus, the kinetic term in our expression essentially summarizes the effect of all the kinetic barriers that need to be traversed in order to get from a given starting state to a given ending state in a fixed amount of time.  Put another way, one more rapidly evolves to states that are separated from one's starting point by fewer high barriers.

The central point in our argument comes from weighing the impact of the remaining two terms, and when doing so, it is important to appreciate what it means to hold the first two terms fixed.  We may imagine a procedure by which the full phase space of the many-particle system of interest is parceled out into macrostate sub-volumes of equal size that are equally kinetically accessible to the starting arrangement.  Quite counterintuitively, although it should generally be the case that the number of such phase space parcels would be extremely small compared to the total number of microscopic arrangements available to the system, it should nonetheless also be the case that the parcels constitute an enormously diverse group of system arrangements with different physical properties.  In particular, we can reasonably expect the members of this group to span a wide range in terms of the values of $\Psi-\Phi$ associated with their formation.  And this means that when we observe a \emph{likely} outcome of this process of stochastic evolution, we should expect it to look to us like it was chosen from a diverse set of options under strong pressure to make the value of $\Psi-\Phi$ more positive.

We have to do some additional work to unpack the physical meaning of this conclusion. For a given starting macrostate $\mathbf{I}$ and ending macrostate $\mathbf{II}$, the quantities $\Psi^{\mathbf{I}\rightarrow\mathbf{II},fwd}$ and $\Phi^{\mathbf{I}\rightarrow\mathbf{II},fwd}$ should generally depend in detail on all the particulars of the system: how the particles in the system interact with each other and with external fields ($H_{sys}$), how the driving field  $\lambda(t)$ changes over time between $t=0$ and $t=\tau$, and what temperature ($\beta$) is maintained in the surrounding heat bath.  More specifically, given these fixed particulars, any choice of $\mathbf{I}$, $\mathbf{II}$ immediately implies a probability distribution $P_{fwd}^{\mathbf{I}\rightarrow\mathbf{II}}(\Delta S_{bath})~dS_{bath}$ for $\Delta S_{bath} =\beta\Delta Q$ that determines the associated values of $\Psi$ and $\Phi$, because any given macroscopic arrangement available to the system will have an average amount of heat that is dissipated during its formation, and also a characteristic amount of fluctuation about that average.

The key question, then, is: what is it about the properties of macrostate $\mathbf{II}$ that would cause it to have a more positive value for $\Psi - \Phi$ than macrostate $\mathbf{III}$? It turns out we can discern different effects as we go term by term.  The value of $\Psi$ corresponds to the mean of the distribution $P_{fwd}^{\mathbf{I}\rightarrow\mathbf{II}}(\Delta S_{bath})~dS_{bath}$, and can be driven up in one of the two ways.  One way of dissipating more heat into the surroundings on average is to go to lower average internal energy; indeed, in an undriven system this is the only possible mechanism for heat dissipation allowed by conservation of energy, which is why the condition of thermal equilibrium must favor states of low energy once internal entropy is held fixed.  However, in driven systems, a second possible mechanism of increasing dissipation comes to the fore: the system can get the time-varying external field to do more \emph{work} on it, which may then be exhausted into the bath as heat.  Moreover, the crucial point here is that not every microscopic arrangement $\mathbf{x}$ available to our system of interest will absorb the same amount of work from the same time-varying field; indeed, the instantaneous work rate of the field at any given moment is $\dot{W} = \partial H_{sys}(\mathbf{x},\lambda(t))/\partial t$, which generally must depend on the value of $\mathbf{x}$ \cite{jarzynski1997nonequilibrium,crooks}.  Thus, to say that the most likely outcomes of driven stochastic evolution should tend to have extremely positive values of $\Psi$ is to suggest they will be states whose formation on average involves passing through configurations that are exceptionally well-suited to the absorption of work from the particular time-varying pattern of $\lambda(t)$.

Viewed in isolation, the tendency in driven systems towards discovery of macrostates that have increased average dissipation $\Psi$ associated with their formation is reminiscent of maximum entropy principles that have been proposed previously \cite{Martyushev2010}.  Here, however, we are forced to consider the countervailing pressure of the fluctuations $\Phi$ on our stochastic evolution.  Quite analogously to the way the effect of energy is offset by that of internal entropy in equilibrium thermodynamics, here the average dissipation is counterbalanced by the dispersion about that average: the more \emph{reliably} one dissipates at a certain level on the way to $\mathbf{II}$, the more likely $\mathbf{II}$ turns out to be as an outcome of the stochastic evolution.  Thus, following macrotrajectories that maximize $\Psi$ alone is not predicted to correspond to a likely outcome for the system because it is possible for the system to explore configurations where both the average dissipation and the associated fluctuation are quite high while their difference is small.  Instead, we must also account for an ``all-things-equal" tendency towards an outcome whose formation is characterized by the suppression of fluctuations and the reliable generation of dissipation over all the likely histories that would have produced that outcome.

Collecting terms together, we can see that a range of different pressures come to bear on how driven systems evolve over time.  On any finite timescale and at any non-zero temperature, considerations of internal entropy and kinetic accessibility play a role, such that what is likely to happen in the system will in part be determined by which states are closest (and thus most kinetically accessible) to the starting state, and also by a general tendency towards internal disorder caused by the randomizing effect of thermal fluctuations. Yet, once these factors or fixed, there remains the question of why the structures we observe form instead of others that might be equally ordered and kinetically stable.  The intriguing suggestion from our analysis of equation (\ref{eq:boltz}) is that the structures that are most likely to form are products of competing pressures to fall to lower internal energy, to further increase heat dissipation through work absorption, and to dissipate that heat as reliably as possible.  In the next section, we will flesh out the consequences of these general physical principles for our understanding of self-organization, adaptation, and biological evolution.

\section{Self-organization and Dissipation-driven Adaptation}

When explaining evolutionary adaptation in the language of biology, we tend to define environments in qualitative terms; organisms are considered to be subject to some external influences that are fixed (such as the terrain and the general climate) and others that oscillate or fluctuate (such as the amount of sunlight or the day-to-day availability of rainwater).  Of course, any of these factors can be couched in more quantitative terms, but it is rare that we consider them in purely physical ones.  Formally, doing so amounts to summarizing the environment experienced by a given collection of particles as consisting of a thermal bath ($T=1/\beta$) and some time-varying external field $\lambda(t)$ that exerts a force on the system's various internal microscopic coordinates.

In the previous section, we demonstrated a general expression for the time-evolution of the probability distribution over macrostates for any such system.  Moreover, we interpreted this equation to be a summary of the different pressures on the organization that emerges in the system.  The first two such pressures were the more obvious ones: all things being equal, we expect our clump of driven matter to tend towards a macrostate that is highly disordered and not separated from our starting arrangement by many high activation barriers.  What remains now is to interpret the remaining pressure to progress to macrostates whose emergence is accompanied by reliably high dissipation, that is, by a large positive value of $\Psi-\Phi$.

How does one reliably increase dissipation?  In answering this question, it is essential that we remember that the system can only dissipate energy into the bath through the system's internal degrees of freedom, because the applied field does not act on the bath.  Thus, one way of enabling more dissipation to happen is by increasing the rate at which positive work is done by the external driving field on the system, and, in this light, we must remember that \emph{not all configurations of the system absorb work from the drive at the same rate}.

This last claim is easiest to appreciate in a simple example, so let us suppose we are driving a damped harmonic oscillator with spring constant $k$, mass $m$, and drag coefficient $b$ using a sinusoidal forcing function $f_{0}\sin[\omega t]$.  It is a basic result in Newtonian mechanics that such a system exhibits a resonance phenomenon, whereby the amplitude of driven oscillation at steady-state exhibits a strong dependence on $\omega$, and is maximized when $\omega= \omega_{r}(k,m,b)$, that is, when the driving frequency is matched to the natural resonance frequency of the oscillator.  Accordingly, the absorbed work and dissipated power also depend on this resonance matching, since added motion gives the drive a greater opportunity to do work via $dW = F~dx$.  Flipping things around, this also means that if we tune $k$ while holding all other parameters fixed, the same forcing function $f_{0}\sin[\omega t]$ will do different amounts of work on the system over time depending on whether the value of $k$ is ``in tune" with the driving environment or not.

The single harmonic oscillator in fact provides the means to make sense of much more complicated systems, since any thermally-coupled many-particle assembly trapped in a local energy minimum has linear-response properties described by a collection of independent damped normal mode oscillations.  More importantly, the spectrum of natural frequencies of oscillation available to the system should be specific to the particular energy minimum in which the system is currently situated.  Thus, the resonant response of our system to a forcing drive of a given frequency should, in general, depend on the particular way in which the system's components are arranged, with the consequence that the amount of work instantaneously absorbed from the same external drive could vary widely depending the system's particular microscopic configuration.

In light of these observations, we can now see equation (\ref{eq:boltz}) as a tug-of-war among competing effects.  In general, thermal fluctuations produce a tendency towards disorder, and the finite time-frame generates a preference for states that are kinetically close to the starting arrangement.  However, it is quite conceivably the case in a given large collection of interacting particles that a very small, special, low-entropy set of arrangements have dynamical response properties that are highly tuned to the driving field, and thus exceptionally well-suited to absorbing external work that can be dissipated as heat.  If the extra dissipation associated with trajectories passing through such states is reliably high enough, it is possible that the likely outcome of the driven stochastic evolution will turn out to be low in internal entropy and in kinetic accessibility to the starting state.  In this event, we would expect the dynamical response properties of the macrostate outcome of this stochastic evolution to exhibit a finely-tuned and detailed relationship to the time-varying field, bearing the mark of the work-absorbing states the system had to pass through during the outcome's emergence.

Adaptation may be construed in various ways, but we at least have succeeded here in developing a language for defining the concept in physical terms.  Not all arrangements of the same collection of matter have the same properties with respect to the flow of energy into and out of the system caused by the application of a particular time-varying field.  What we have now is general theoretical argument for why driven systems should tend preferentially to traverse organized states that have exceptional properties in terms of the reliable absorption and dissipation of energy for the particular driving environment they experience.  Thus, we are justified in saying that the driven system progresses to states that are better adapted to the drive over time, and it is clear that this emergent adaptation should be accompanied by organization into particular special shapes whose physical properties would appear to us to be tailored to the features of the particular environment. 

It must be stressed here that we have not made \emph{any} reference to self-replication and natural selection in motivating this physical argument.  The adaptation we predict is expected to take place independent of whether or not there is anything in the system that can copy itself and pass heritable traits to a range of similar progeny.  This curious result demands further inquiry, for while we may be confident that the general theoretical considerations on which it is based are valid, it remains quite a mystery how a driven system ``learns" to absorb energy better from its particular driving environment if not through successive rounds of reproduction and natural selection.  In the next section, we will investigate the connection between reliable dissipation and the likely outcomes of stochastic evolution in a simple, concrete model.  Using the concepts we develop in that setting, we will be able to suggest a more intuitive explanation for our prediction of spontaneous adaptation in strongly driven many-particle systems.
\section{Dissipation and Drift in Time-varying Energy Landscapes}

To build intuition for how external entropy-production through heat dissipation is connected to the dynamical behavior of nonequilibrium systems, it is helpful to consider a simple, stochastic model of a single particle hopping in a discrete landscape of energy levels.  Each state $x_{i}$ has a particular energy $E_{i}$, and each pair of states $x_{i}$ and $x_{j}$ are separated by an activation barrier of energy $B_{ij}\equiv B_{ji} > \max(E_{i},E_{j})$.  To introduce the idea of a thermal bath of temperature $T\equiv 1/\beta$ in a way that is consistent with our underlying assumptions about time-reversibility, we make the Arrhenius assumption that, when our particle is in state $x_{i}$, it always has a constant probability rate $r_{i\rightarrow j}$ of stochastically hopping to state $x_{j}$ that is given by \cite{gardiner} 
\begin{equation}
r_{i\rightarrow j} = r^{0}_{ij}\exp\left[-\beta(B_{ij}-E_{i})\right]
\end{equation}
Here, $ r^{0}_{ij}= r^{0}_{ji}$ is a time constant specific to the pair of states that remains unaffected when we change $E_{i}$, $E_{j}$, or $B_{ij}$.  By assuming transition probability rates to take this form, we are guaranteed that, so long as the energies of states and barriers do not change with time, the probability distribution for the location of our hopping particle must at steady-state converge on a Boltzmann distribution $p_{s}(x_{i})\propto \exp[-\beta E_{i}]$ that will obey the detailed balance condition $p_{s}(x_{i})/p_{s}({x_{j}}) = r_{j\rightarrow i}/r_{i\rightarrow j} $  required by time-reversal symmetry.

To introduce the idea of an external driving field, it is only necessary to allow the energies $E_{i}(t)$ and $B_{ij}(t)$ to be arbitrary functions of time.  Work is done on the system whenever the energy $E_{i}$ changes while the particle is located at $x_{i}$, and heat $\Delta Q = E_{i}(t)-E_{j}(t)$ is exhausted into the surrounding thermal bath whenever the particle hops from $x_{i}$ to $x_{j}$ at time $t$.  

There many different phenomena in stochastic dynamics that could be realized in a class of models as general as the one we have just described.  However, our specific interest in this case is to illustrate the meaning of equation (\ref{eq:boltz}) in concrete terms, so as to clarify how fluctuation and dissipation affect the flow of probability during stochastic evolution.  An important detail in the interpretation of that equation was that the average external entropy production $\Psi$ and the fluctuations about that average $\Phi$ only come to the fore in determining the relative likelihood of events in our stochastic process if we compare outcomes of the same internal entropy ($\ln\Omega$) and return probability $\pi^{rev}_{\tau,\lambda(t)}$.  In the single-particle hopping scenario we are now considering, the question of $\ln\Omega$ is moot because each microstate $x_{i}$ is assumed to have the same internal entropy ($\ln 1 = 0$).  

Fixing the return probability, however, requires considering more a specific scenario, for it is not immediately intuitive what it means to compare two possible outcomes of a stochastic process that are equally likely to run backwards to their common starting point.  In the system at hand, however, we may contrive an example that has this property.  Consider a landscape composed of only three states $x_{1}$, $x_{2}$, $x_{3}$ arranged in a row so that $r_{12}^{0} = r_{23}^{0}=r>0$ and $r^{0}_{13} = 0$ (Figure 3).  All three states are initially assumed to have the same energy $E_{i} = 0$ and to be interlaced with two barriers of equal height $B_{12} = B_{23}=\Delta E>0$.

If the single particle is placed at $x_{2}$ at $t=0$, then, in the absence of any external drives, it is obviously equally likely to hop to $x_{1}$ on ``the left" or to $x_{3}$ on ``the right" in some short time $\tau\ll 1/r$ since $r_{2\rightarrow 1} = r_{2\rightarrow 3} = r e^{-\beta\Delta E}$ .  Moreover, the probability of returning to $x_{2}$ in time $\tau$ after making that first jump is also the same for particles at $x_{1}$ and $x_{3}$.  We may choose, however, to drive the system at high frequency so that $E_{1}(t) = -\Delta E\cos(\omega t)/2$ and  $B_{12}(t) = \Delta E-\Delta E\cos(\omega t)/2$, while $E_{2}$, $E_{3}$, and $B_{23}$ remain constant in time and $\omega\gg1/\tau\gg r$.

 \begin{figure}
 \includegraphics[width=\columnwidth]{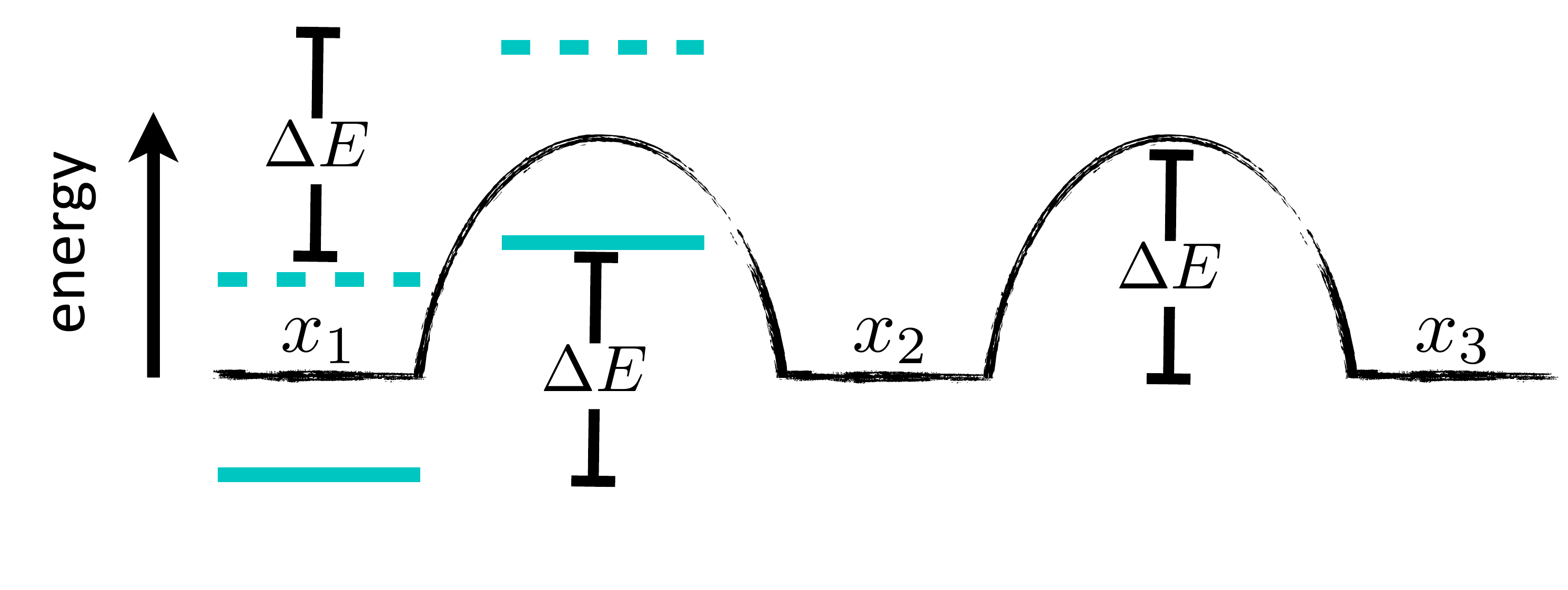}
 \caption{Three separate microstates initially of equal energy $E=0$ are arranged on a line so that stochastic transitions may take place between adjacent locations separated by activation barriers of height $\Delta E$.  When state $x_{1}$ and the barrier separating it from $x_{2}$ are driven, such that their energies oscillate in time, a preference for $x_{1}$ over $x_{3}$ as a destination for finite time evolution from $x_{2}$ develops.  This statistical drift is necessarily accompanied by the conversion of absorbed work into dissipated heat in the surroundings.  The energies of the driven states are drawn at their extremal values in blue, with dashed lines and solid lines respectively in phase with each other. }
 \end{figure}

The total effect of this particular scenario of driving is several-fold.  First of all, it is clear that, as desired, $r_{1\rightarrow 2}= r_{3\rightarrow 2} = r e^{-\beta\Delta E}$, meaning that states $x_{1}$ and $x_{3}$ constitute a pair of outcomes for evolution from $x_{2}$ that have equal return probability.  This condition is achieved by a contrivance that causes $B_{12}(t) - E_{1}(t) = \Delta E = B_{23}-E_{3}$ to remain constant over time, so that return jumps from either state always must surmount a barrier of fixed height.  

What is also apparent, however, is that by fixing return probability in this driven landscape, we have necessarily also introduced a strong correlation between heat dissipation and drift.  The rightward hopping rate $r_{2\rightarrow 3} = re^{-\beta \Delta E}$ does not change with time.  In contrast, the leftward hopping rate gets an extra boost from the oscillation of the barrier energy $B_{12}(t)$, so that
\begin{equation}
r_{2\rightarrow 1}(t) = r \exp\left[-\beta(\Delta E-\Delta E\cos[\omega t]/2)\right]
\end{equation}
So long as $\beta\Delta E\gg 1$, leftward hopping events are much more likely to happen during the part of the drive cycle when the hopping rate is maximal because of the exponential rate-enhancement that comes from the downswing of $B_{12}(t)$.  At this maximum, the ratio of the leftward and rightward rates is
\begin{equation}
\frac{r_{2\rightarrow 1}^{max}}{r_{2\rightarrow 3}}= \exp\left[\beta\Delta E/2\right] 
\end{equation}
indicating a strong bias towards landing at $x_{1}$ rather than at $x_{3}$ after initially being placed at $x_{2}$.  One may confirm this bias by averaging each hopping rate over a whole drive period $2\pi/\omega$, obtaining $\overline{r_{2\rightarrow 1}} = r e^{-\beta\Delta E}I_{0}(\beta\Delta E/2)>\overline{r_{2\rightarrow 3}}=r e^{-\beta\Delta E}$.

More crucially, however, we can now appreciate that the fact of this bias towards one outcome over the other requires that there reliably be extra heat dissipation that accompanies motion in the likely direction.  During moments in the drive cycle when transits from $x_{2}$ to $x_{3}$ are likely (such as at $t=0$), our requirement of fixed return probability has guaranteed that $E_{1}$ will be at its minimum value, so that a hop to the left must reliably be accompanied by a positive amount of dissipative external entropy production $\Psi_{2\rightarrow 1} =\beta\langle \Delta Q\rangle_{2\rightarrow 1} = \beta \Delta E/2$.  However, hops to the right will not dissipate at all since $E_{3}=E_{2}=0$ do not change with time, so $\Psi_{2\rightarrow 3} =0$.  Thus, we can see that reliable entropy production in the surroundings and concerted drift towards a more likely outcome of stochastic evolution are two sides of the same coin once we have restricted ourselves to comparing outcomes of constant return probability.  Referring back to equation (\ref{eq:boltz}), we can see we have recovered the predicted relationship
\begin{equation}
\frac{r_{2\rightarrow 1}^{max}}{r_{2\rightarrow 3}}= \exp\left[\Delta\Psi\right] 
\end{equation}

Using similar reasoning, we may also consider the effect of the fluctuations $\Phi$ on drift.  The above discussion has amounted to an argument for why scenarios involving drift in a particular direction must also exhibit elevated external dissipation (that is, more positive $\Psi$), yet it is clearly the case that not all heat evolution in driven systems is accompanied by such statistically irreversible flow of probability density.  Indeed, using the Crooks relation for microstate-to-microstate transitions in such a discrete model, one may easily show that when $\exp[-\beta\Delta Q]$ is averaged over all stochastic trajectories that both start and end in the same state $x_{i}$, it is generally the case that $\Psi_{i\rightarrow i} = \Phi_{i\rightarrow i}\geq 0$, reminding us that much of the heat generated in a driven system can result from futile cycles that go around in circles rather than drifting in some new direction in phase space \cite{hatano}. Thus, some ways of making $\Psi$ more positive also bring about an increase of $\Phi$ in lockstep that cancels any contribution to the flow of probability in equation (\ref{eq:boltz}).  Put another way, high fluctuations in entropy production lessen the degree to which drift and dissipation are correlated.

To see this effect of fluctuations in the case of our hopping particle, we can specialize to a slightly different different scenario (Figure 4).  Considering now a pair of states $x_{1}$ and $x_{2}$, we can initialize the particle in state $x_{2}$ with constant energy $E_{2}=0$ and suppose there are two distinct paths that connect the states.  The first path has a barrier height $B_{12}(t) = \Delta E + \Delta E\cos(\omega t)/2$, whereas the second path has a barrier of height $B_{12}^{*}(t) = \Delta E - \Delta E\cos(\omega t)/2$, and both barriers are governed by the same time constant $r$. The energy of state $x_{1}$, meanwhile is assumed to vary with time as $E_{1}(t) =    \Delta E\cos(\omega t)/2.$ 

 \begin{figure}
 \includegraphics[width=\columnwidth]{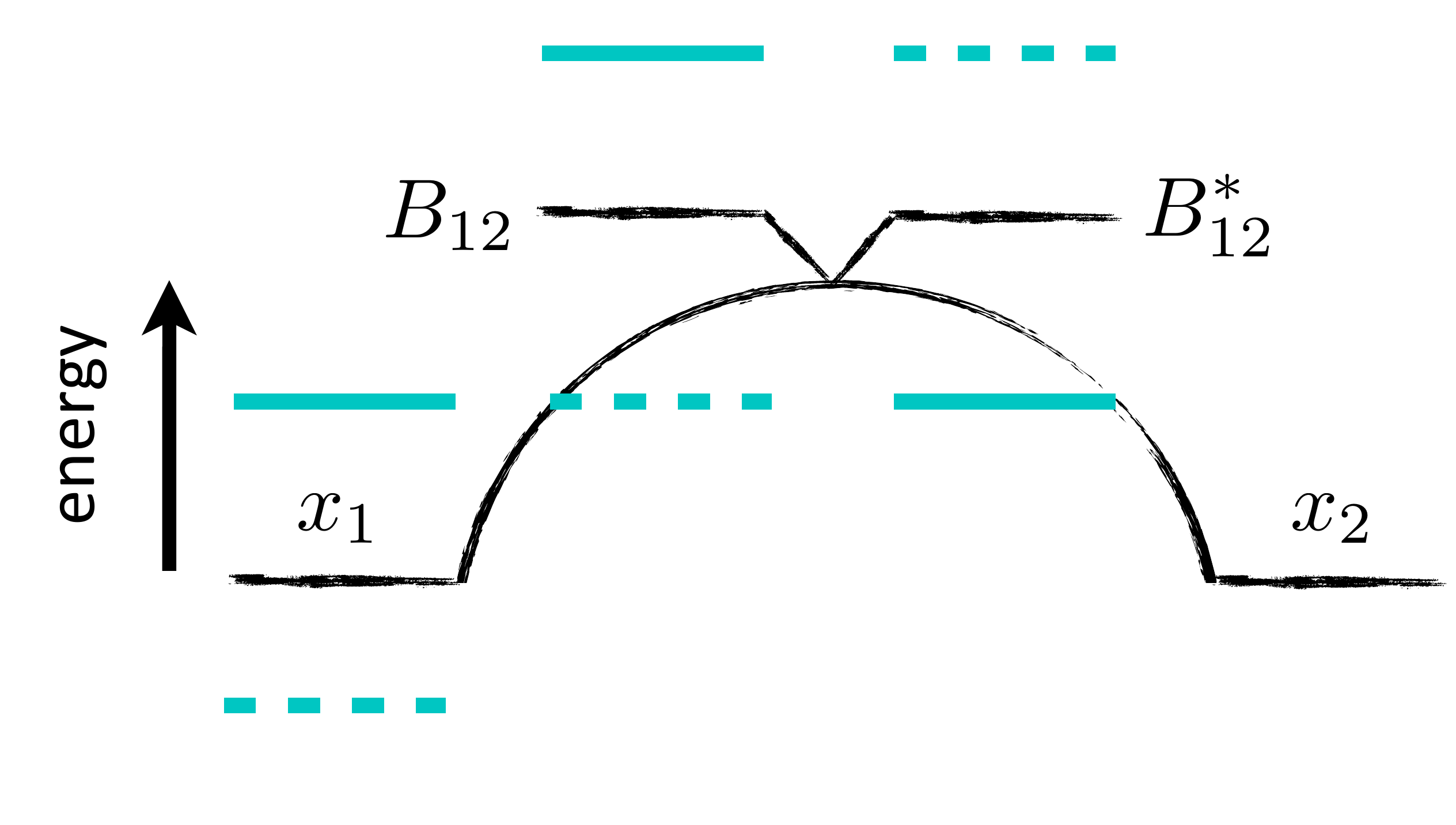}
 \caption{In this driven scenario, two states $x_{1}$ and $x_{2}$ are separated by two different barriers representing two different possible paths for transiting between the states.  Because the barriers are driven out of phase with each other, the return probability and average dissipation for transitions from $x_{2}$ to $x_{1}$ are independent of $\Delta E$, the choice of which should still affect the drift rate from $x_{2}$ to $x_{1}$ as well as the fluctuations $\Phi$ for the pair of paths.  As in Figure 3, the energies of the driven states are drawn at their extremal values in blue, with dashed lines and solid lines respectively in phase with each other.  }
 \end{figure}

In a system thus described, we can straightforwardly compute all of the relevant quantities if we assume we are in a regime where $\beta\Delta E\gg 1$ and $r \ll 1/\tau\ll \omega$, so that transitions are rare events that take place during the moments in the drive cycle of their maximum likelihood.  In this case, it is easy to see that the particle is equally likely to first jump to $x_{1}$ via either barrier, since $\max(r_{2\rightarrow 1}(t))= \max(r_{2\rightarrow 1}^{*}(t))=r\exp[-\beta\Delta E/2]$.  Interestingly, however, the mean external entropy production (that is, the heat dissipation) averaged over the weights of these paths should be $\Psi =\beta (\Delta E/2 - \Delta E/2)/2 = 0$, due to the fact that events on one path happen in phase with the rise of $E_{1}$ and events on the other path happen $\pi$ radians out of phase with it.  Moreover, in the limit of $\Delta E$ we have chosen, the total return probability $r\tau \textnormal{const.} $ is effectively insensitive to the exact value of the barrier height, since the return journey will always be made via $B_{21}^{*}$ and not $B_{21}$. However, in this same regime, changes in $\Delta E$ \emph{strongly} affect both the forward transition rate and the fluctuations in entropy production; specifically the transition rate is $\ln\overline{r_{2\rightarrow 1}^{tot}}=\ln[2re^{-\beta\Delta E}I_{0}(\beta\Delta E/2)]\simeq-\beta \Delta E/2$ and the fluctuations are $\Phi = 0-(-\ln\langle e^{-\Delta S}\rangle) =\ln\cosh[\beta\Delta E/2]\simeq \beta \Delta E/2 $.  Thus, for this particular case we may write what we again should have expected from equation (\ref{eq:boltz}):
\begin{equation}
\ln\overline{r_{2\rightarrow 1}^{tot}} \simeq - \Phi
\end{equation}
from which it is apparent that, as fluctuations in entropy production rise with return probability and average dissipation already held fixed, the forward probability rate must fall sharply.  The origin of this effect lies in the diversity of paths connecting states, and the resulting possibility that the drive can sometimes \emph{help} you return to your starting point via a different path than it pushed you along when your journey first began.

The simple examples explored here help provide some intuition for how dissipation is sometimes, but not always coupled to statistically irreversible drift in driven systems.  What remains for us now is to connect these intuitions to the dynamics of complex, many-particle arrangements that are capable of exhibiting emergent organization in which we might recognize behaviors reminiscent of evolutionary adaptation.

To begin making this connection, we may first recognize that if we finely discretize the phase space of a macroscopic collection of atoms or molecules, we may think of the whole microtrajectory of the system as that of a single particle performing stochastic hops along a lattice of possible locations in an extremely high dimensional space.  In this mapping, the number of atomic degrees of freedom becomes the number of independent directions in which the hopping particle can move, and, just as before, the energies of the states and the transition rates between may be thought of as functions that vary with time in a manner consistent with Newtonian time-reversibility.

In qualitative terms, we can recognize a few consequences of such a mapping.  First of all, for a macroscopic system with a typical inter-atomic Hamiltonian at a temperature at which many types of chemical bonds are able to persist for a range of timescales, the undriven energy landscape experienced by our `hopping particle' should be thought of as fairly rugged, since kinetic traps abound.  The effect of the drive, then, is to substantially alter the transition rates connecting various pairs of states that are not rapidly inter-accessible from thermal fluctuations alone.

Secondly, we should also expect that with so many independent directions in which the `particle' can diffuse, it ought to behave like it is confronted by a certain amount of quenched randomness in its surrounding energy landscape.  At the same time, one expects this randomness to contain many correlations that reflect the fact that a particular location for the `particle' corresponds to a particular arrangement for all the matter in system, so that neighboring states will share many of the same physical properties.

The importance of these correlations comes into relief when we start to drive the system, and energies of states and barriers start to change over time in ways that are specific to the type of drive and the types of atoms in the system.  For, while it is true that we do not always expect any given direction in phase space to look like the archetypal drift scenario proposed in the previous section, it is quite reasonable to suppose that some of the $\sim 10^{25}$ directions in the phase space of the system of interest should have a structure that happens to mimic the pattern needed to pump probability density irreversibly in one direction, at least to some extent.  Moreover, it is also plausible that the correlation in physical properties that must exist between neighboring regions of phase space should mean that some regions much more effectively absorb external work and power irreversible drift than do others.  Taken together, these suggestions force us to consider the possibility that concerted drift into phase space volumes particularly suited to work absorption from the external drive is an outcome we should expect in general for sufficiently many kinetically trapped degrees of freedom driven far enough from equilibrium

\section{Discussion}

In this article, we have sought to provide the beginnings of a physical account of adaptation. Starting from very general assumptions, we have argued for the emergence of `well-adapted' structures in systems of many interacting particles driven far from thermal equilibrium by a time-varying external field.  We first motivated this argument from a more abstract theoretical perspective by writing down the generalization of the Helmholtz free energy for the finite-time stochastic evolution of arbitrarily driven nonequilibrium macrostates.  From this point of view, we were able to make the case that when highly ordered, kinetically stable structures form far from equilibrium, it must be because they achieved reliably high levels of work absorption and dissipation during their process of formation.  It is appealing to describe such structures as well-adapted to their environment because of the special matching to external drives they must exhibit in order to achieve such an exceptional history of thermodynamic flux.  

Subsequently, in order to make the reasons for the predicted effects more intuitive and concrete, we specialized to a class of simple systems obeying Arrhenius kinetics and demonstrated how dissipation and fluctuation each affect the flow of probability density during stochastic evolution.  These considerations established more clearly that, while statistically irreversible drift must be reliably accompanied by extra heat evolution, there are other `high fluctuation' mechanisms of heat evolution that do not imply drift. We went on to make a plausibility argument for why these intuitions from simple models should still apply in the more general setting where many interacting degrees of freedom are driven at once;  in short, we suggested that in a very high-dimensional, rugged energy landscape, there will always be directions in phase space for any given drive that happen to use work absorption to power irreversible drift.  After that drift has occurred, one expects to see a structure that appears `better-adapted,' since its history was one of achieving a special matching to the surrounding driving fields that allowed it to absorb extra work. 

The implications of the physical notion of adaptation put forward here are potentially quite broad.  In the biological context, it is possible to understand the origins of evolutionary adaptation in terms of the survival and reproduction of individual self-replicators capable of passing on their traits to future generations.  Here, however, we have made no assumption involving self-replication in motivating the proposed mechanism of physical adaptation and `learned' work absorption.  Nonetheless, there are several points that should be made in connection to the Darwinian picture.  To start, we may note that if our system of interest turns out to be made of self-replicators, then the Darwinian account of adaptation and the thermodynamic one given here become one and the same.  Previously \cite{England2013}, authors of this work have shown that self-replication's statistical irreversibility imposes the requirement that each self-copying event must be accompanied on average by a minimum positive amount of dissipation in the surroundings.  Thus, taking over the future through exponential growth in a Darwinian competition may be seen as a process whose likeliest outcomes are generated through reliably high dissipation.  

Moreover, the notion of generating reliable dissipation through sustained statistically irreversible drift in phase space in general presents a bit of a puzzle to us, since this imposes the requirement that a structure currently well-suited to work absorption must be using that work energy to \emph{change} its shape in order for the dissipation to be \emph{reliably} high.  One might therefore worry that the change in shape could disrupt the flow of work powering the change. Self-replicating forms of organization present a very appealing resolution of this puzzle, since they provide a means to change the overall structure of the system (by making a copy) while preserving the pattern of organization in the system that was so particularly well-suited to the absorption of work.  Thus, it may not be so surprising that the most striking examples of adaptation that we have observed come from biological self-replicators.

At the same time, perhaps the most interesting implication of the thermodynamic picture of adaptation we offer here is that there may be many examples of `well-adapted' structures that did not have parents.  It has certainly long been known that an endless variety of far-from-equilibrium many-particle Newtonian systems are capable of exhibiting self-organization phenomena in which strikingly patterned structures emerge in the presence of dissipative external drives \cite{prigogine1971biological}.  Whether in sand dunes or snow flakes, in hurricanes or in spiral bundles of protein filaments and motors \cite{schaller2010polar,sanchez2012spontaneous}, the nonequilibrium world offers many test cases for the general hypothesis that organized, kinetically stable structures emerge and persist because their formation is reliably accompanied by extra work absorption and dissipation.  While the story in each of these cases must be different in many of its details, we may speculate that a thermodynamic commonality would be revealed to underly all of them if the right physical observables were tracked and compared.

Guided by these theoretical considerations, we recently undertook to demonstrate the predicted phenomenon in a simple simulation framework that tracks the vibration spectrum of a sinusoidally driven toy chemical mixture over time.  The main result of the study (which will be published in a separate article, rather than shown here) is that we do indeed see emergent `adaptive' resonance our system, such that our choice of frequency for an external driving field determines the location of the peak in the resonance spectrum for a particle mixture the evolves stochastically in the presence of such an environmental drive.  This finding turns out to be highly suggestive of the results of recent experiments performed on silver nanorods assembling in the presence of light fields of different colors \cite{Ito2013}, which similarly `learned' to match their surface plasmon resonance to the frequency of the driving field.  Thus, we are encouraged to explore further with related models, spurred on by the intriguing possibility that life-like behavior in nonequilibrium systems may turn out to be surprisingly common, now that we have begun learning how to look for its physical signatures.

\bibliography{evol2014.bbl}

\end{document}